# The near room-temperature upsurge of electrical resistivity in Lu-H-N is not superconductivity, but a metal-to-poor-conductor transition

Di Peng[1,2,3], Qiaoshi Zeng[1,4,*], Fujun Lan[1], Zhenfang Xing[1,5], Yang Ding[1], Ho-kwang Mao[1,4,*]

1. Center for High Pressure Science and Technology Advanced Research, Shanghai 201203, China
2. Key Laboratory of Materials Physics, Institute of Solid State Physics, HFIPS, Chinese Academy of Sciences, Hefei 230031, China
3. University of Science and Technology of China, Hefei 230026, China
4. Shanghai Key Laboratory of Material Frontiers Research in Extreme Environments (MFree), Shanghai Advanced Research in Physical Sciences (SHARPS), Shanghai 201203, China
5. State Key Laboratory of Superhard Materials, Institute of Physics, Jilin University, Changchun 130012, China

* E-mail: zengqs@hpstar.ac.cn, or maohk@hpstar.ac.cn

Since the discovery of superconductivity in mercury at ~4 K in 1911, searching for materials with superconductivity at higher temperatures towards practical conditions has been a primary enduring goal. The recent report of room-temperature superconductivity at near-ambient pressure in nitrogen-doped lutetium hydride (Lu-H-N) by Dasenbrock-Gammon *et al.*[1] (hereafter referred as D-G) seems a great step approaching the ultimate goal. Specifically, they claimed "evidence of superconductivity on Lu-H-N with a maximum $T_c$ of 294 K at 1 GPa." However, the failure to observe the drastic temperature-dependent resistance change above 200 K in high-pressure synthesized Lu-H-N compounds, a prerequisite for superconductivity, by researchers worldwide in all independent follow-up studies[2-11] casts a heavy shadow on the authenticity of the claims. The sober questions are: what is the sample that produces the sharp resistance jump near room temperature? What are the reasons for the non-reproducibility of others who follow the D-G method of synthesis and the inscrutable low success rate (35%) in synthesizing the right sample even for the authors in Ref.[1]? What causes the observed sharp resistance jump? Here, with a well-controlled experiment protocol, we repeatedly reproduced the near room-temperature sudden change of electrical resistance in the Lu-H-N sample, and we could quantitatively compare its behavior with the initial pure Lu in a normal metallic state. These results enable us to scrutinize the origin for the near-room temperature sharp resistance change, which is attributed to a metal-to-poor-conductor transition rather than superconductivity.

We develop a reliable protocol to minimize the extensively concerned possible extrinsic influences of the sample[1,12] by starting the reaction from a piece of pure Lu foil loaded with $H_2/N_2$ gas mixture and conduct in-situ Van der Pauw four-probe resistance measurement[13] to monitor the entire reaction process in a diamond anvil cell (DAC) under different pressures and temperatures using the Physical Property Measurement System (PPMS, Quantum Design). Temperature dependent resistance of the sample (including the initial pure Lu metal and the later on reacted Lu-H-N) at target pressures with different reaction time can be obtained without extra sample change, manipulation, and exposure to air (More experimental details can be found in the Method). Microphotographs of Lu foil with $H_2/N_2$ gas mixture and a Van der Pauw four-probe circuit at ~10 GPa and room temperature (295 K) are shown in the insets of Fig.1. The initial resistance-temperature curve (blue squares in Fig.1) shows a positive temperature coefficient of resistance and residual resistance below ~15 K, featuring a typical normal metallic state of Lu. However, after 5-days holding at ~10 GPa and 295 K, an apparent reaction between Lu and $H_2/N_2$ gas mixture can be evidenced by the sample color change (from silver to dark blue), slightly increased sample size, and the shrunk sample chamber (a considerable amount of $H_2/N_2$ gas has been consumed). Indeed, the temperature dependent resistance curve of the reacted sample (red dots in Fig.1) becomes quite different from the initial one of the pure Lu metal. A sharp upsurge of resistance emerges at ~250 K during warming. Although here we confirmed the sharp resistance transition reported by D-G [1], it should be emphasized that with the measurement of the initial pure Lu metal as the baseline, we can conclude that the entire resistance curve, above and below the sharp change at ~250 K of the reacted sample, are higher than that of the initial pure Lu metal. In other words, the sharp change is caused by the sample forming a much worse metal above 250 K. Its resistance at temperatures below 250 K is still higher than that of pure Lu, contradicting to the superconducting claim. The experiment has been repeated and reproduced without failure. The sharp resistance change has also been observed at releasing pressure to 5 and 2 GPa.

The temperature dependence of resistance of the reacted sample below ~230 K is linear and basically parallel to that of the initial Lu metal. Therefore, it is not justified to treat the low-temperature linear part as an extra meaningless system background, as did by the authors in Ref. [1] to derive zero-resistance. The hallmark and the most desirable property of a superconductor, is its zero electrical resistance, which enables an electric current to flow without any energy dissipation. In this aspect, the absence of zero-resistance and instead upsurge of resistance rules out the possibility of a superconducting transition in the Lu-H-N system and makes this Lu-H-N sample interesting only as a poor metal or semiconductor, not a superconductor.

In addition, the sample color change from blue to pink was claimed in Ref. [1] to be a critical feature associated with the sudden resistance change. Therefore, many groups worldwide devoted tremendous efforts searching for the pink crystal with superconductivity[5,8,11,14]. Although they synthesized the blue Lu-H-N sample using various methods and observed the blue-to-pink transition during compression, they all failed to reproduce the near-room temperature sharp resistance transition[2,5,6,8-11]. The authors in Ref. [1] did not show sample microphotographs with clear details of how the four platinum (Pt) leads connected to the sample and how the temperature dependent resistance gradually evolves with sample color changes. Our results indicate that it is unnecessary to link the blue-to-pink color change with a sharp resistance transition, consistent with the previous conclusions on the non-superconducting pink Lu-H-N samples[2]. The microphotographs in a very recent paper [15] (its Figs. 1 and S1) using the same sample provided by the authors in Ref. [1] reveal that the D-G sample showing the sharp resistance change is not a pure substance, but actually an uneven composite mixture, with undefined textual and ratio of metallic and insulating phases showing shining metallic, gray, black, and pink color under reflective lighting, rather than a pure pink phase as shown in the inset of Fig. 2a in Ref. [1]. The low success rate of D-G *et al.*[1] and no-success of others[2,5,6,8-11] are consistent with expected resistance of the random multiple metal-insulator phase mixture. Clearly, zero resistance cannot be defined in such mixture, and a controlled baseline such as the initial pure Lu resistance before reaction in the present work is essential for consideration of superconductivity.

In summary, by taking the approach of in situ resistance measurements of pure Lu foil during its reaction with $H_2/N_2$ gas mixture at broader pressure and temperature space, we confirmed that the abrupt resistance transition near room temperature could be reproduced in the dark blue Lu-H-N sample but not at the experimental conditions provided in Ref. [1]. The increase of the sample resistance after reaction compared with that of the initial pure Lu metal rules out attributing the observed sudden resistance change to any superconducting transition. Hopefully, with the concrete and reproducible results in this work, the science community could invest precious time and resources in the right direction towards addressing many questions raised on the Lu-H-N system other than the ostensible near room-temperature, room-pressure superconductivity.

## Methods

**Sample synthesis and characterization.** The Lu-H-N sample was prepared by a reaction between pure Lu foil and $H_2/N_2$ gas mixture at a certain temperature and pressure. A high-purity block of Lu (99.9%, Alfa Aesar) and $H_2/N_2$ gas mixture (volume ratio is 99:1) were used as the raw materials for the synthesis of Lu-H-N. Lu foil with a thickness of ~10 μm was obtained from the block of metal Lu by cutting and forging. Then, square samples with the side length of ~80 μm were cut from the thin foil. High-pressure chemical reactions and resistance measurements were conducted in a customized beryllium-copper (Be-Cu) alloy DAC using a non-magnetic rhenium gasket. The diamond anvil culet size is ~300 μm. Cubic boron nitride with epoxy resin was filled in laser-drilled holes of a pre-indented rhenium gasket with a diameter of ~300 μm and a thickness of ~30 μm as an insulating material for electrical measurements. Ruby balls are loaded along the Lu sample as a pressure calibrant. Four platinum electrodes were arranged symmetrically on the Lu sample to monitor the resistance variation during the chemical reaction. $H_2/N_2$ gas mixture was then loaded into the Be-Cu DAC using a gas loading system, which was vacuumed and pre-purged twice with the $H_2/N_2$ gas mixture. The Be-Cu DAC with four platinum electrodes were connected to the PPMS instrument for in situ resistance measurements on the sample during reaction at different target pressure and temperature conditions.

**Acknowledgments** The authors acknowledge financial support from Shanghai Science and Technology Committee, China (No. 22JC1410300) and Shanghai Key Laboratory of Novel Extreme Condition Materials, China (No. 22dz2260800). The authors thank Haiyun Shu for his kind help with gas loading and Prof. Liling Sun for her helpful discussion.

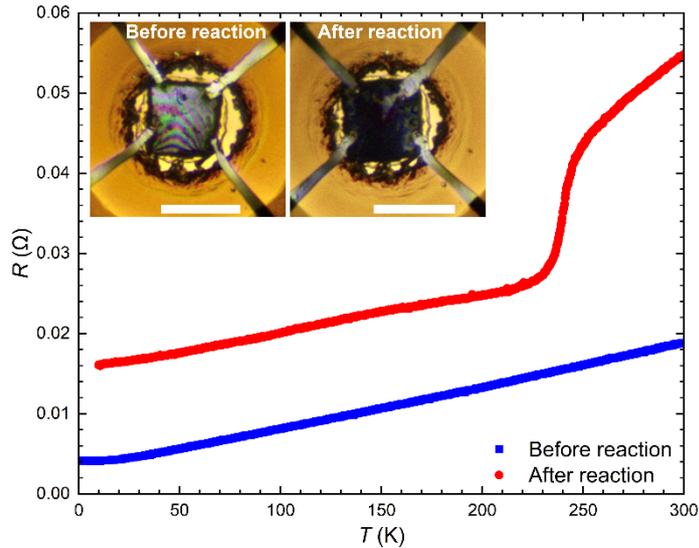

**Fig. 1 | In situ electric resistance measurements for the Lu foil sample loaded with $H_2/N_2$ gas mixture in a DAC at 10 GPa and 295 K.** The insets are the optical microphotographs of the Lu foil sample just loaded (left) and after a 5-days reaction (right) with four Pt electrodes and $H_2/N_2$ gas mixture in a DAC at 10 GPa and 295 K. The metallic luster surface of the Lu changes into dark blue after a 5-days reaction. The colorful fringes are due to light interference between the gap from the sample to the diamond anvil surfaces, which becomes invisible after reaction mainly due to much less light reflection by the dark sample surface. The obvious consumption of $H_2/N_2$ gas mixture can be reflected by sample chamber shrinkage after a 5-days reaction. Sample size also slightly increases after the reaction. The blue and red curves are temperature dependent raw resistance values during warming from 2 K (10 K) to 300 K for the initial pure Lu metal before reaction and the same sample after reaction for 5 days at 10 GPa and 295 K, respectively.

# References

1. Dasenbrock-Gammon, N. *et al.* Evidence of near-ambient superconductivity in a N-doped lutetium hydride. *Nature* **615**, 244-250 (2023).
2. Ming, X. *et al.* Absence of near-ambient superconductivity in $LuH_{2\pm x}N_y$. *Nature*, 10.1038/s41586-41023-06162-w (2023).
3. Cai, S. *et al.* No evidence of superconductivity in a compressed sample prepared from lutetium foil and $H_2/N_2$ gas mixture. *Matter Radiat. Extrem.* **8** (2023).
4. Ball, P. Superconductivity feels the heat. *Nat. Mater.* **22**, 404 (2023).
5. Shan, P. *et al.* Pressure-Induced Color Change in the Lutetium Dihydride $LuH_2$. *Chin. Phys. Lett.* **40** (2023).
6. Sun, Y., Zhang, F., Wu, S., Antropov, V. & Ho, K.-M. Effect of nitrogen doping and pressure on the stability of cubic $LuH_3$. *arXiv:2303.14034v1* (2023).
7. Liu, M. *et al.* On parent structures of near-ambient nitrogen-doped lutetium hydride superconductor. *arXiv:2303.06554v1* (2023).
8. Zhang, Y.-J. *et al.* Pressure induced color change and evolution of metallic behavior in nitrogen-doped lutetium hydride. *Sci. China Phys. Mech.* **66**, 287411 (2023).
9. Zhang, S. *et al.* Electronic and magnetic properties of Lu and $LuH_2$. *AIP Adv.* **13**, 065117 (2023).
10. Hilleke, K. P. *et al.* Structure, stability and superconductivity of N-doped lutetium hydrides at kbar pressures. *arXiv:2303.15622v1* (2023).
11. Xing, X., Wang, C., Yu, L. & Xu, J. Observation of non-superconducting phase changes in $LuH_{2\pm x}N_y$. *arXiv:2303.17587* (2023).
12. Wang, N. *et al.* Percolation-induced resistivity drop in cold-pressed $LuH_2$. *arXiv:2304.00558* (2023).
13. van der Pauw, L. J. A Method of Measuring Specific Resistivity and Hall Effect of Discs of Arbitrary Shape. *Philips Res. Rep.* **13**, 1-9.
14. Tao, X., Yang, A., Yang, S., Quan, Y. & Zhang, P. Leading components and pressure-induced color changes in N-doped lutetium hydride. *Sci. Bull.* (2023).
15. Salke, N. P., Mark, A. C., Ahart, M. & Hemley, R. J. Evidence for Near Ambient Superconductivity in the Lu-N-H System. *arXiv*, 2306.06301 (2023).